\documentstyle{article}
\newcommand{\dx}{\partial _x}
\newcommand{\dy}{\partial _y}

\title{Integrable systems with quadratic nonlinearity in Fourier space}
\author{V.G. Marikhin \footnote{e-mail: mvg@itp.ac.ru}\\
L.D. Landau Institute for Theoretical Physics,\\
Kosygina st. 2, 117333 Moscow, Russia}
\begin{document}
\maketitle
\thispagestyle{empty}
\begin{abstract}
The Lax pair representation in Fourier space is used to obtain a list of integrable
scalar evolutionary equations with quadratic nonlinearity. The
famous systems of this type such as KdV, intermediate long-wave
equation (ILW), Camassa-Holm and A. Degasperis systems are
represented in this list. Some new systems are obtained as well. The
generalizations on two-dimensional and discrete systems are
discussed.
\end{abstract}
\vspace{7mm}

An important class of the integrable evolution systems is a class of systems
with quadratic nonlinearity. Traditionally, the evolution integrable systems
are classified by power of dispersion law.
We don't assume a concrete form of dispersion law beforehand.
Moreover, we consider dispersionless systems also.

We shall consider the systems of following type
\begin{equation}\label{kdvn}
\frac{d}{dt}u_{q}(t)=\omega (q)
u_q(t)+\sum\limits_{p_1+p_2=q}w(p_1,p_2)u_{p_1}(t)u_{p_2}(t)
\end{equation}
The summation in (\ref{kdvn}) has a symbolic meaning. One can
consider a finite set of values of "momentum" $p$ - this case
corresponds to the finite-dimensional systems. The other case
$p\in\Re$ corresponds to the one-dimensional evolution equations.

In fact, one can apply the inverse Fourier transform
(we will omit an imaginary unit)
to equation (\ref{kdvn}) to obtain
the systems in a coordinate space.

For example,  one can
substitute $\omega(k)=k^3,\; w(p,q)=p+q$ to the equation (\ref{kdvn})
and apply the inverse Fourier transform
to obtain the KdV equation. Another example:
the intermediate long-wave water (ILW) equation corresponds to
a choice $\omega(k)=k^2\frac{1+e^{hk}}{1-e^{hk}},\; w(p,q)=p+q$
in equation (\ref{kdvn}).

One of the most effective tools used in the theory of integrable
systems is Lax pair representation.
One can try to obtain Lax representation of (\ref{kdvn}) in the following form
\begin{equation}\label{lax1}
L_t=[A,L],\quad L=\alpha+\sum\limits_p U^p u_p,\quad A=\beta+\sum\limits_p V^p
u_p,\quad [\alpha,\beta]=0
\end{equation}
where $\alpha,\,\beta,\,U^p,\,V^p$ - some constant operators.
It is easy to see that (\ref{lax1}) is Lax pair for (\ref{kdvn}) if
and only if
\begin{equation}\label{laxUV}
[V^p,U^q]+[V^q,U^p]=2w(p,q)U^{p+q},\quad
[\beta,U^p]+[V^p,\alpha]=\omega (p) U^p
\end{equation}
One can use a following matrix representation for the operators in
Fourier space
\begin{equation}\label{abUV}
\alpha_{kk'}=\alpha (k)\delta_{k,k'},\;\beta_{kk'}=\beta (k)\delta_{k,k'},
\quad U^p_{kk'}=l(k,p)\delta_{k,k'+p},\;V^p_{kk'}=a(k,p)\delta_{k,k'+p}
\end{equation}
One can substitute (\ref{abUV}) to (\ref{laxUV})  to obtain
\begin{equation}\label{arho}
a(k,q)l(k-q,p)+a(k,p)l(k-p,q)-a(k-q,p)l(k,q)-a(k-p,q)l(k,p)=2w(p,q)l(k,p+q)
\end{equation}
and
\begin{equation}\label{albe}
\lbrack\alpha(k)-\alpha(k-p)\rbrack a(k,p)=\lbrack\beta(k)-\beta(k-p)-\omega(p)\rbrack l(k,p)
\end{equation}

It is easy to obtain a particular solution of (\ref{albe}).
\begin{equation}\label{salbe}
\alpha(k)=l(k,0),\;\beta(k)=a(k,0),\;\omega(k)=2w(k,0).
\end{equation}
This solution is compatible with (\ref{arho}), but is
trivial because it corresponds to a shift $u_p(t)\rightarrow
u_p(t)+\delta_{p,0}$ in (\ref{kdvn}).

We shall consider  $l(k,p)=\frac{1}{\rho(k)}$ case now, then equation (\ref{arho})
transforms to
\begin{equation}\label{ar}
a(k,q)\frac{\rho(k)}{\rho(k-q)}+a(k,p)\frac{\rho(k)}{\rho(k-p)}-
a(k-q,p)-a(k-p,q)=2w(p,q)
\end{equation}
It is natural to differentiate (\ref{ar}) on $p$ and on $q$ to
obtain
\begin{equation}\label{ag}
a(k,p)=kA(p)+b(p)+c(k),\;2w(p,q)=pA(q)+qA(p)+B(p)+B(q),\;
b(-p)=b(p)+pA(p)+B(p)
\end{equation}
where $A(p)=A(-p),\; B(p)=-B(-p),\;\omega(p)=-\omega(-p).$

One can substitute (\ref{ag}) to (\ref{ar}) to obtain a
functional equation
\begin{equation}\label{U}
U(x,y)=U(y,x),\quad U(x,y)=\rho(x)[xA(x-y)+b(x-y)+c(x)]
\end{equation}
that defines a whole class of integrable dispersionless $(\omega_p=0)$ systems
with quadratic nonlinearity because one can choose $\alpha=0,\;\beta=0$ in this case.

One can rewrite (\ref{albe}) to obtain another functional equation
\begin{equation}\label{disp}
\beta(x)-\beta(y)-\omega(x-y)=(\alpha(x)-\alpha(y))U(x,y)
\end{equation}
where $U(x,y)$ is some solution of (\ref{U}).

The general solution of (\ref{disp}) defines a class of integrable
systems (\ref{kdvn}) with nonzero dispersion $(\omega_p\neq 0).$
Evidently, this solution doesn't exist for all $U(x,y)$ from
(\ref{U}).
\section{Classification of dispersionless systems}
It is possible to obtain the general solution of (\ref{U}).

{\bf The first step} is to apply operators $D=\dx+\dy,\;D^2\;D^3$ to
equation (\ref{U}) to obtain three more equations - it is
important that $D\,f(x-y)\equiv 0.$
Therefore, equation (\ref{U}) transforms to
\begin{equation}\label{det4}
W_D(x\rho(x)-y\rho(y),\;\rho(x),\;\rho(y),\;\rho(x)c(x)-\rho(y)c(y))=0
\end{equation}
where $D-$Wronskian is defined by formula
\begin{equation}\label{Dw}
W_D(f_1(x,y),\,f_2(x,y),..,f_N(x,y))=\det(W_i^j),\quad
W_i^j=D^{j-1}f_i(x,y),\quad i,j=1,2,..,N
\end{equation}
{\bf The second step} is to apply $y\rightarrow x$ limit to
equation (\ref{det4}) to obtain an {\bf necessary} condition
\begin{equation}\label{nec1}
c(x)=\mu_1\frac{R(x)}{\rho(x)}+\mu_2 x+\mu_3,\quad
\lambda_1 \rho''(x)+\lambda_2 x\rho'(x)+\lambda_3 \rho'(x) +\lambda_4
\rho(x)=0,\quad R'(x)=\rho(x)
\end{equation}
There are two cases - case $\lambda_2=0$ is "trivial" - any
solution of (\ref{nec1}) corresponds to a solution of
(\ref{det4}):

$\rho(x)=c_1\,e^{h_1 x}+c_2\,e^{h_2 x}$ or $\rho(x)=(x-a)e^{h x}$
or $\rho(x)=e^{h x}$ or $\rho(x)=x$ or $\rho(x)=1.$

In $\lambda_2\neq 0$ case one can substitute
$x\rho(x)=-\frac{1}{\lambda_2}(\lambda_1 \rho'(x)+(\lambda_3-\lambda_2) \rho(x) +\lambda_4
R(x))$ to (\ref{det4}), to take the $y\rightarrow x$ limit as well
to obtain the second necessary condition.
\begin{equation}\label{nec2}
\nu_4\rho^{IV}(x)+\nu_2\rho''(x)+\nu_1\rho'(x)+\nu_0=0
\end{equation}

In this case we have only three solutions different from $\lambda_2=0$
case: $\rho(x)=(x-a)(x-a-b)(x-a+b)$ or $\rho(x)=(x-a)(x-b)$ -
Degasperis \cite{deg1},\cite{deg} and Camassa-Holm \cite{ch} systems respectively
and $\rho(x)=1/x$ case.

So we have obtain following classification of dispersionless systems with
quadratic nonlinearity:
\begin{enumerate}
\item $\rho(k)=1$

In this class one can found the hydrodynamic type systems
\begin{equation}\label{hy}
u_t=mu_x,\quad m_p=A(p)u_p,\quad
L_{kk'}=u_{k-k'},\;A_{kk'}=\frac{1}{2}(k+k')\,A(k-k')u_{k-k'}.
\end{equation}
\item $\rho(k)\neq 1,\; A(k)=0$
\begin{itemize}
\item The system from this class have a general form in a
$\rho(x)=e^{hx}$ case
\begin{equation}\label{a00}
u_t=m\,u,\quad m_p=B(p)u_p,\quad
L_{kk'}=e^{-hk}u_{k-k'},\;A_{kk'}=b(k-k')u_{k-k'},\quad B(p)=b(p)(e^{hp}-1).
\end{equation}
It is interesting that a choice $b(x)=\frac{1}{1+e^{hx}}$ corresponds to
Hilbert-Hopf equation $m_t=L(\partial_x) m^2,\; L(k)=th\frac{kh}{2}$
where $(u_k=m_k(e^{hk}+1)).$

\item The case $\rho(x)=(x-a)\,e^{hx}$
corresponds to
\begin{equation}\label{xex}
u_t=m\,u,\quad m_p=B(p)u_p,\quad
L_{kk'}=(k-a)^{-1}e^{-hx}u_{k-k'},\;A_{kk'}=(\frac{1}{k-a}+\frac{e^{h(k'-k)}-1}{k-k'})u_{k-k'}.
\end{equation}
where $\quad B(p)=\frac{1}{2p}(2-e^{hp}-e^{-hp})$

\item Two-exponent case $\rho(x)=e^{h_1x}+e^{h_2x}$
corresponds to
\begin{equation}\label{ex2}
u_t=m\,u,\quad m_p=B(p)u_p,\quad B(p)=b(-p)-b(p),\quad
b(x)=\frac{b_1(1-e^{h_1x})-b_2(1-e^{h_2x})}{e^{h_1x}-e^{h_2x}}
\end{equation}
where $
L_{kk'}=(a_1e^{h_1k}+a_2e^{h_2k})^{-1}u_{k-k'},\;
A_{kk'}=(b(k-k')+c(k))u_{k-k'},\;
c(k)=\frac{b_1a_1e^{h_1k}+b_2a_2e^{h_2k}}{a_1e^{h_1k}+a_2e^{h_2k}}.$
\end{itemize}
Note that the systems (\ref{xex}) and (\ref{ex2}) have a second
Lax pair representation (\ref{a00}).
\item $\rho(k)\neq 1,\; A(k)\neq 0$
\begin{itemize}
\item First example in this class is a system
with $\rho(x)=x^2-1/4:$
Lax pair  has a form
$$L_{kk'}=\frac{1}{k^2-1/4}u_{k-k'},\quad
A_{kk'}=\frac{1}{2}(\frac{k-3k'}{1-(k-k')^2}+\frac{k}{k^2-1/4})u_{k-k'},$$
that corresponds to Camassa-Holm \cite{ch} equation
\begin{equation}\label{ch}
u_t=2f_xu+fu_x,\quad u=\frac{1}{2}(f_{xx}-f)
\end{equation}

\item Second example is a system with $\rho(x)=x(x^2-1):$
Lax pair in this case has a form
$$L_{kk'}=\frac{1}{k(k^2-1)}u_{k-k'},\quad
A_{kk'}=(\frac{k-2k'}{1-(k-k')^2}+\frac{k}{k^2-1})u_{k-k'}.$$
that corresponds to A. Degasperis system \cite{deg1},\cite{deg}
\begin{equation}\label{dega}
u_t=3f_xu+fu_x,\quad u=\frac{1}{2}(f_{xx}-f)
\end{equation}

The other cases are
\item Systems with $\rho(x)=x$ and $A_{kk'}=k'A(k-k')$
have a form
\begin{equation}\label{rx}
u_t=\dx (mu),\quad m_p=A(p)u_p
\end{equation}
\item Systems with $\rho(x)=\frac{1}{x}$ and $A_{kk'}=kA(k-k')$
have a form
\begin{equation}\label{r1x}
u_t=mu_x-m_xu,\quad m_p=A(p)u_p
\end{equation}
\end{itemize}
\end{enumerate}
\section{The systems with dispersion}
One can try to solve equation (\ref{disp}) starting from the general solution of
(\ref{U}) to obtain a list of systems with dispersion.
One can verify that there are no solutions of (\ref{disp})
for systems (\ref{ch}) and (\ref{dega}).
So the statement is: It is impossible to obtain a generalization of
Camassa-Holm (\ref{ch}) and A. Degasperis (\ref{dega})
on the systems with dispersion. This statement is valid for the systems
(\ref{xex}) and (\ref{ex2}) as well.

The rest of possibilities are systems
with $\rho(x)=1, \;\rho(x)=e^{hx}$ or $\rho(x)=x.$

\begin{enumerate}
\item In a case $\rho(x)=1$ it is easy to obtain
$U(x,y)=\frac{1}{2}(x+y)A(x-y)+d(x-y),$ where $A(x),\, d(x)$ are
even functions. One can substitute given $U(x,y)$ to (\ref{disp}),
apply the operator $D$ to equation to obtain a necessary condition
\begin{equation}\label{rh1}
W_D(\alpha(x)-\alpha(y),(x+y)(\alpha(x)-\alpha(y)),\beta(x)-\beta(y),1)=0
\end{equation}
The method of solution of equation (\ref{rh1}) is similar to that
of solution of (\ref{det4}):
Limit $x\rightarrow y$ gives $\beta'(x)=c\,x\alpha'(x).$
Substitution $\beta(x)$ to (\ref{rh1}) and taking $x\rightarrow y$ limit
give a necessary condition
$\alpha''(x)+c_1x\alpha'(x)+c_2\alpha'(x)+c_3=0.$
The only possible choice to fulfill (\ref{rh1}) is $\alpha(x)=x^2,\beta(x)=4x^3$
(one can choose $c=6$).

It is easy to obtain
$\omega(x)=x^3$ in this case, that corresponds to KdV
equation $u_t=u_{xxx}+6uu_x$ with
\begin{equation}\label{kdvla}
L_{kk'}=k^2\delta_{kk'}+u_{k-k'},\quad
A_{kk'}=4k^3\delta_{kk'}+3(k+k')u_{k-k'}.
\end{equation}
\item In $\rho(x)=e^{hx}$ case one have
$U(x,y)=e^{\frac{1}{2}(x+y)}d(x-y),$ where $d(x)$ is even function.

The necessary condition in this case
\begin{equation}\label{rh2}
W_D((\alpha(x)-\alpha(y))e^{\frac{1}{2}(x+y)},\beta(x)-\beta(y),1)=0
\end{equation}
One have $\beta'(x)=c_1\alpha'(x)e^{hx}+c_2.$
The only non-trivial solution of (\ref{rh2})
$\alpha(x)=x\,e^{-hx},\beta(x)=x^2$ gives intermediate long-wave
(ILW) equation
\begin{equation}\label{ilw}
u_t=\Gamma(u_{xx})-2uu_x,\quad \Gamma(p)=\coth (\frac{ph}{2})
\end{equation}
with Lax pair
\begin{equation}\label{ilwla}
L_{kk'}=k\,e^{-hk}\delta_{kk'}+e^{-hk}u_{k-k'},\quad
A_{kk'}=k^2\delta_{kk'}+2\frac{k-k'}{1-e^{h(k-k')}}u_{k-k'}
\end{equation}
\item
In $\rho(x)=x$ case we just obtain
alternative Lax pair for KdV equation:
\begin{equation}\label{kdval}
L=\delta_{kk'}(\frac{k}{2}+\frac{\mu}{k})+\frac{1}{k}u_{k-k'},\quad
A=\delta_{kk'}(\frac{k^3}{6}-\mu k)+k'u_{k-k'},\quad L_t=[A,L]
\iff u_{p,t}=\frac{p^3}{6}u_p+p\sum\limits_q u_qu_{p-q}.
\end{equation}
\end{enumerate}
\section{Generalizations}
It is possible to extend our approach to high-dimensional and
multi-component systems.
Consider two-dimensional scalar systems with quadratic
nonlinearity. Unlike an one-dimensional case, the
"wave" variables $k,p$ and $q$ in the functional equations (\ref{arho})
and (\ref{albe}) are vectors with two components.
To obtain the functional equations in this case one can use following trick:
One can write $\vec{k}=(k_x,k_y)=(k,\lambda),$ where $k$ is an
argument of the functional equations and $\lambda$ is parameter.
For example, one can rewrite equation (\ref{U}) as follows
\begin{equation}\begin{array}{l}\label{U2}
\rho(x,\lambda)[xA(x-y,\lambda-\mu)+\lambda
A_2(x-y,\lambda-\mu)+b(x-y,\lambda-\mu)+c(x,\lambda)]=\\
\rho(y,\mu)[yA(y-x,\mu-\lambda)+\mu
A_2(y-x,\mu-\lambda)+b(y-x,\mu-\lambda)+c(y,\mu)]
\end{array}\end{equation}
The method of solution of (\ref{U2}) is analogous to
one-dimensional ones. One can use notations $\rho(x)=\rho(x,\lambda),\,c(x)=c(x,\lambda),\;
\tilde{\rho}(y)=\rho(y,\mu),\,\tilde{c}(y)=c(y,\mu)$ to obtain
\begin{equation}\label{det42}
W_D(x\rho(x)-y\tilde{\rho}(y),\;\rho(x),\;\tilde{\rho}(y),\;
\rho(x)c(x)-\tilde{\rho}(y)\tilde{c}(y))=0
\end{equation}
in this case (compare with (\ref{det4})).

Next step is to take $\mu\rightarrow\lambda$ limit in equation
(\ref{det42}) to obtain equation (\ref{det4}) exactly!
Note that one can't apply this limit to functional equation
(\ref{U2}) because possible divergence in the functions $A,A_2,b.$

Then the only possibility to obtain two-dimensional systems with
quadratic nonlinearity is to start from one-dimensional systems
and to try to extend some solutions of (\ref{U}) to (\ref{U2}).
We have no full classification in two-dimensional case at the
moment, but a couple of examples are in agree with our approach:

First example is a famous Kadomtsev-Petvishvili (KP) equation
$$u_t=u_{xxx}+3\theta^{-1}(u_y)+6uu_x,\;\theta^{-1}=\partial_y\partial_x^{-1}$$
has a Lax pair in Fourier space
\begin{equation}\label{kp}
l_{kk'}=(k_y+k_x^2)\delta_{kk'}+u_{k-k'},\quad
a_{kk'}=4k_x^3\delta_{kk'}+3(k_x+k'_x-\frac{k_y-k'_y}{k_x-k'_x})u_{k-k'}
\end{equation}
that is an agreement with the general form of Lax pair (\ref{ag})
in our method.

Second example is Veselov-Novikov (VN) equation
\begin{equation}\label{VN}
u_t=u_{xxx}+u_{yyy}+u\theta(u_x)+u_x\theta(u)+u\theta^{-1}(u_y)+u_y\theta^{-1}(u),
\;\theta=\partial_x\partial_y^{-1}
\end{equation}
One can consider VN equation (\ref{VN}) as a sum of two symmetries
$u_t=u_{t+}+u_{t-}$ where
$u_{t+}=u_{xxx}+u\theta(u_x)+u_x\theta(u)$ and
$u_{t-}=u_{yyy}+u\theta^{-1}(u_y)+u_y\theta^{-1}(u).$

Lax pair in Fourier space for "+" symmetry has a form
\begin{equation}\label{vn}
L_{kk'}=3k_y\delta_{kk'}+\frac{1}{k_x}u_{k-k'},\quad
A_{kk'}=k_x^3\delta_{kk'}+k'_x\frac{k_x-k'_x}{k_y-k'_y}u_{k-k'}
\end{equation}

One can introduce new operators:
\begin{equation}\label{trop}
L_1=L,\;L_2=L[x\leftrightarrow y],\;A_1=A,\;A_2=A[x\leftrightarrow
y],\quad L_{i,t_i}=[L_i,A_i],\quad {\bf L}=L_1+L_2,\;{\bf A}=A_1+A_2,\;
\end{equation}
It is easy to obtain the Manakov's triada representation for Veselov-Novikov
equation (\ref{VN})
\begin{equation}\label{tri}
{\bf L}_t=[{\bf A},{\bf L}]+B{\bf L},\quad B=A_1^{(1)}+A_2^{(2)}-A_1-A_2
\end{equation}
where $A_{1,kk'}^{(1)}=k_xA_{1,kk'}\frac{1}{k'_x},\;A_{2,kk'}^{(2)}=k_yA_{2,kk'}\frac{1}{k'_y}
\Rightarrow B_{kk'}=\frac{(k_x-k'_x)^2}{k_y-k'_y}+\frac{(k_y-k'_y)^2}{k_x-k'_x}$

A case of multi-component systems corresponds to a choice
$k\rightarrow (k,n)$ where $k\in\Re$ and $ n=1,2,..,N.$
A classification of those systems is  a subject of forthcoming
papers, but in a case of two-component systems we have two
interesting classes:

First example is so called Toda-type systems
$$z_{q,t}=-\omega (q)z_q+\gamma (q)p_q+\sum\limits_s\Gamma^{(1)}(s,q-s)z_{s}z_{q-s},
\quad p_{q,t}=\omega (q)p_q+\sum\limits_s\Gamma^{(2)}(s,q-s)z_{s}p_{q-s}$$
with Lax pair
\begin{equation}\label{lato}
L=\alpha+\sum\limits_qU^qz_q+\sum\limits_qW^qp_q\quad
A=\beta+\sum\limits_qV^qz_q,\quad [\alpha,\beta]=0
\end{equation}
Note that Toda system has Lax pair representation in coordinate
space
\begin{equation}\label{tla}
L_t=[A,L],\quad L=\partial+z-\lambda+p\partial^{-1},\quad
A=\partial^2+2\partial\,z \iff z_t=\partial [-z'+z^2+2p],
\quad p_t=\partial [p'+2pz].
\end{equation}
This dynamical system can be transformed to nonlinear
Schr\"odinger equation $(p=\psi^*\psi,\; z=-\partial_x\log(\psi))$
and B\~acklund transformation of this system have a form of Toda
lattice (see \cite{inll} for more details).
It is easy to see that Fourier transform of (\ref{tla}) has a form
(\ref{lato}).

The other example is a system with 3-wave interaction
\begin{equation}\begin{array}{l}\label{3wave}
\;\;\; i\,a_{p,t}=\omega (p) a_p +\sum\limits_q W^{(1)}(p,q)a_{p+q}a_{q}^*+
\sum\limits_q W^{(2)}(p,q)a_qa_{p-q}\\
-i\,a^*_{p,t}=\omega (p) a^*_p +\sum\limits_q \bar{W}^{(1)}(p,q)a^*_{p+q}a_{q}+
\sum\limits_q \bar{W}^{(2)}(p,q)a^*_qa^*_{p-q}
\end{array}\end{equation}

The problem is to find all possible functions $W^{(1)},W^{(2)}$ and
$\omega$ in case when system (\ref{3wave}) is integrable. This question
was discussed in \cite{ZaSc} by the use of perturbative approach.

Lax pair in this case has a form
\begin{equation}\label{3w}
L=\alpha+\sum\limits_{p>0}U^pa_p+\sum\limits_{p>0}(U^{p})^+a^*_p,\quad
A=\beta+\sum\limits_{p>0}V^pa_p+\sum\limits_{p>0}(V^{p})^+a^*_p ,\quad
[\alpha,\beta]=0.
\end{equation}
One can derive a system of functional equations as in scalar
systems case, but a general solution of these equations is unknown as yet.
It is interesting that system (\ref{3wave}) is equivalent to
scalar system (\ref{kdvn}) in a "real" case:
$$u_p=\theta(p)a_p+\theta(-p)a_{-p}^*,\;u_p^*=u_{-p},
\quad W^{(1)}(p,q)=w(q,p-q),\;W^{(2)}(p,q)=w(p+q,-q).$$

In summary, we have use a Fourier representation of Lax pairs to classify scalar
integrable systems with quadratic nonlinearity.
The similar (but not the same) approach is so called symbolic method.
This method was applied to theory of integrable systems by Gelfand and Dickii
\cite{gd} and was improved in the recent papers
\cite{sund},\cite{nvs1}.
The common feature and main advantage of both methods is that all coefficients of
equations in Fourier space
are the functions (not operators!) of wave numbers $k.$
Because this fact, all necessary conditions can be formulated as
the functional equations. The irony of it is that in some cases it is better to solve
functional equations rather then differential ones.

This work was supported by INTAS No. 99-01782 and Russian foundation
for Basic Research No. 01-01-00874.

\end{document}